\documentclass[letter]{aa}
\usepackage{graphicx}
\usepackage{txfonts}
\usepackage[]{natbib}

\def \src {IGR~J11215--5952}

\def \hcm {\hbox {\ifmmode $ atom cm$^{-2}\else atom cm$^{-2}$\fi}}

\def \arcsec {\hbox{$^{\prime\prime}$}}

\def \ATel {ATel}

\defcitealias{SidoliPM2006}{Paper~I}
\begin{document}
   \title{Swift/XRT observes the fifth outburst of the periodic Supergiant
Fast X--ray Transient IGR J11215--5952 }

   \author{P. Romano\inst{1,2} \and L.\ Sidoli\inst{3} \and   V.\ Mangano\inst{4} \and 
	S.\ Mereghetti\inst{3} \and  G.\ Cusumano\inst{4} }
   \offprints{P.\ Romano, patrizia.romano@brera.inaf.it}
   \institute{INAF, Osservatorio Astronomico di Brera, Via E.\ Bianchi 46, 
	I-23807 Merate (LC), Italy 
    \and Universit\`a{} degli Studi di Milano, Bicocca, Piazza delle Scienze 3, 
	I-20126 Milano, Italy  
    \and INAF, Istituto di Astrofisica Spaziale e Fisica Cosmica, 
	Via E.\ Bassini 15,   I-20133 Milano,  Italy          
    \and INAF, Istituto di Astrofisica Spaziale e Fisica Cosmica, 
	Via U.\ La Malfa 153, I-90146 Palermo, Italy 
             }
   \date{Received 1 March 2007/Accepted 4 April}
\abstract
{IGR J11215-5952 is a hard X--ray transient source discovered in April 2005 with 
INTEGRAL and a confirmed member  of the new class of High Mass X--ray Binaries, 
the Supergiant Fast X--ray Transients (SFXTs).
Archival INTEGRAL data and RXTE observations showed that the outbursts occur
with a periodicity of $\sim$330 days.
Thus, \src\ is the first SFXT displaying periodic outbursts, possibly related
to the orbital period.
}
   {We performed a Target of Opportunity observation with Swift
with the main aim of monitoring the source behaviour around
the time of the fifth  outburst, expected on 2007 Feb 9.
}
{The source field was observed with Swift twice a day (2ks/day)
starting from 4th February, 2007, until the fifth outburst,
and then for $\sim 5$\,ks a day afterwards,
during a monitoring campaign that lasted 23 days for a total on-source 
exposure of $\sim73$\,ks. This is the most complete monitoring campaign of 
an outburst from a SFXT.
}
{The spectrum during the brightest flares is well described by an
absorbed power law with a photon index of 1 and $N_{\rm H} \sim 1 
\times 10^{22}$ cm$^{-2}$. A  1--10\,keV peak luminosity of $\sim $10$^{36}$~erg~s$^{-1}$ 
was derived (assuming 6.2~kpc, the distance of the optical counterpart).}
{These Swift observations are a unique data-set for an outburst of a SFXT,
thanks to the combination of sensitivity and time coverage, and they allowed a 
study of \src\ from outburst onset to almost quiescence. 
We find that the accretion phase lasts longer than previously thought on the basis 
of lower sensitivity instruments observing only the brightest flares.  
The observed phenomenology is consistent with a smoothly increasing flux
triggered at the periastron passage in a wide eccentric orbit with
many flares superimposed, possibly due to episodic or inhomogeneous
accretion.
}
\keywords{X-rays: stars: individual: \src  }
\authorrunning {P.\ Romano et al.}
\titlerunning {The fifth outburst of \src\ observed by Swift}
\maketitle
\section{Introduction}

	\begin{figure*}
	 \sidecaption
 	 	\includegraphics[width=14cm,height=14cm]{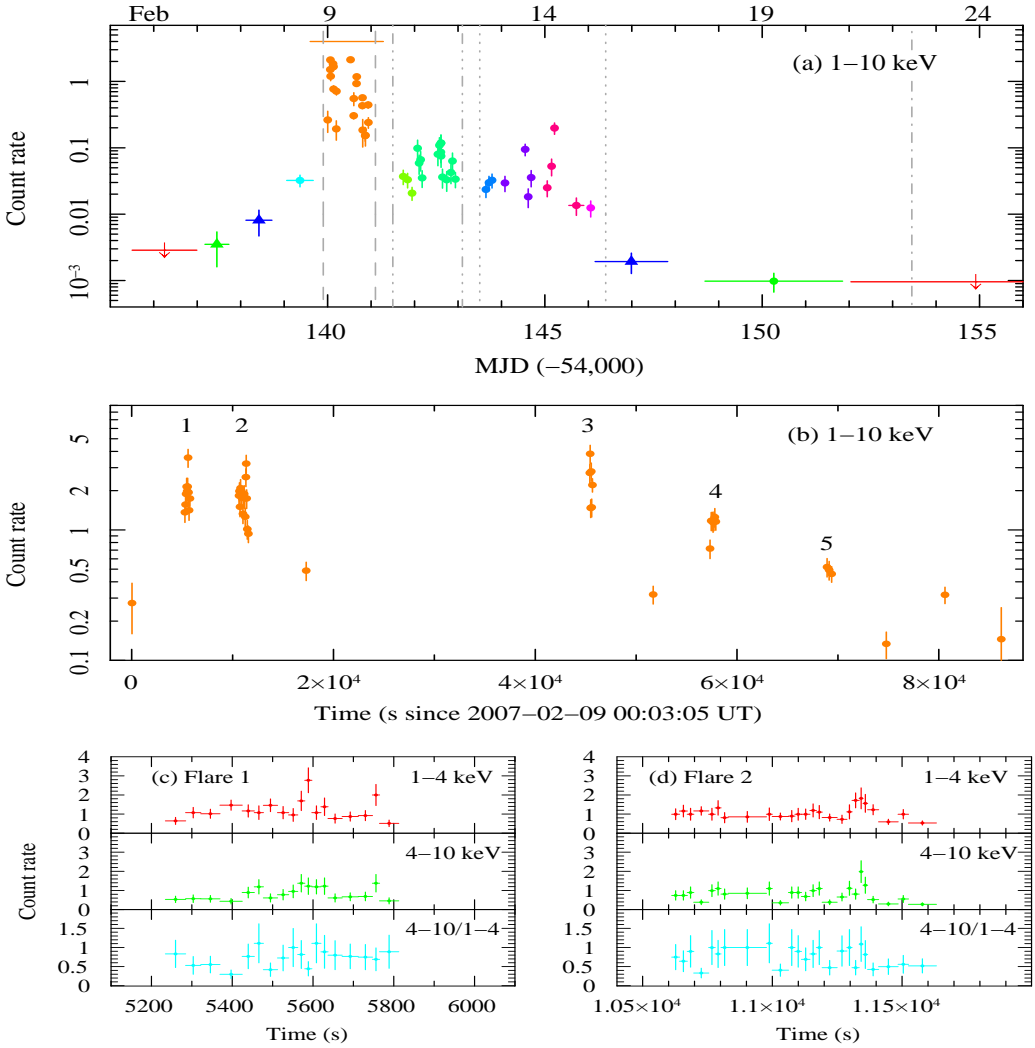}
		\caption{XRT light curves, corrected for pile-up, PSF losses, vignetting and 
		background-subtracted. 
		{\bf a)} 1--10\,keV light curve for the whole campaign.
		Different colours denote different observations (Table~\ref{igr112:tab:alldata}),
		and points before Feb 6 (MJD 54,137) and after Feb 15 (MJD 54,146) 
		are drawn from the sum of several observations. 
		Filled circles are full detections (S/N$>$3), 
		triangles marginal detections ($2<$S/N$<3$), 
		while downward-pointing arrows are 3-$\sigma$ upper limits. 
		The vertical lines mark our time selections for spectroscopic analysis
		(observation 6, observations 7--8, observations 9--12, end of observation 18). 
		The horizontal line marks the region shown enlarged in panel b). 
		{\bf b)} Detail of observation 6, with a binning that allowed to achieve a S/N in excess of 6. 
		The numbers mark the 5 flares on which we performed spectroscopic analysis. 
		{\bf c)} Detail of flare 1, showing the 1--4\,keV, 4--10\,keV count rates 
			(top and middle panel) and the hardness ratio 4--10/1--4 (bottom).  
			The data were rebinned in order to have at least 20 counts per	
			bin in both the 1--4\,keV and 4--10\,keV band. 
		{\bf d)} Same as c), for flare 2. 
		}
 		\label{igr112:fig:lcv}
	\end{figure*}
The hard X--ray transient \object{IGR~J11215--5952} was discovered with the
INTEGRAL satellite during an outburst in  April 2005 \citep{Lubinski2005}
and was associated with \object{HD~306414} 
\citep{Negueruela2005b}, a B1Ia  supergiant 
located at a distance of 6.2~kpc \citep{Masetti2006}. 
The short duration of the outburst
together with the likely optical counterpart
suggested that \src\ could be a new member of the 
class of Supergiant Fast X-ray Transients 
(SFXTs; \citealt{Negueruela2005a}). 
Analysing archival INTEGRAL observations of the source field,
\citet*[hereafter Paper I]{SidoliPM2006} 
discovered two previously unnoticed outbursts (in July 2003 and in May 2004) 
which demonstrate the recurrent nature of this transient and suggest a
possible periodicity of $\sim$330 days.
This periodicity was confirmed by the detection of the fourth outburst
from \src\ with RossiXTE/PCA  on 2006 March 16--17, 
329 days after the third outburst \citep{Smith2006a}. 
The RXTE/PCA observations showed strong flux variability
and a hard spectrum (power-law photon index  
of $1.7\pm{0.2}$ in the range 2.5--15 keV) as well as 
a possible pulse period of  $\sim$195\,s \citep{Smith2006b}. 
The periodicity was confirmed with RXTE observations of the latest outburst,
yielding $P=186.78\pm0.3$\,s \citep{Swank2007:atel999}.
Follow-up observations with Swift/XRT refined the source position
and confirmed the association with HD~306414 \citep{Steeghs2006}. 
A hard power-law with a high energy cut-off around 15~keV is a good fit to the
spectra observed with INTEGRAL (Paper~I). 
For the distance of 6.2 kpc, the peak fluxes of the outbursts
correspond to a luminosity of $\sim 3 \times$10$^{36}$~erg~s$^{-1}$ (5--100~keV).
All these findings confirmed \src\ as a member of the 
class of the SFXTs, and  the first object of
this class of High Mass X--ray Binaries displaying periodic outbursts. 

Predicting a fifth outburst for 2007 Feb 9, we obtained a 
Target of Opportunity (ToO) observing campaign with Swift, which commenced 
on Feb 4. The source started showing renewed activity on Feb 8 
\citep{Romano2007:atel994} 
and underwent a powerful outburst on Feb 9 
\citep{Mangano2007:atel995,Mangano2007:atel996,Sidoli2007:atel997,Swank2007:atel999}.
This paper presents our observations of \src\ 
and it is organized as follows. 
In Sect.~\ref{igr112:dataredu} we describe our observations and data reduction; 
in Sect.~\ref{igr112:dataanal} we describe our  spatial, timing and spectral data analysis. 
Finally, in Sect.~\ref{igr112:discussion} we discuss our findings and draw our conclusions.

  \section{Observations and Data Reduction\label{igr112:dataredu}}

Table~\ref{igr112:tab:alldata} reports the log of the Swift/XRT observations 
used for this work. 
Thanks to Swift's fast-slewing and flexible observing scheduling, the ToO observations 
started on 2007 Feb 4 with 2\,ks per day evenly spread throughout the day to 
maximize the chances of detection of the outburst onset, and were increased to 
5\,ks afterwards
for a total of 23 days and a total on-source exposure of $\sim 73$\,ks.	
We also retrieved from the Swift Archive the data from a 643\,s ToO  
performed on 2006 Mar 20 during the fourth 
outburst of this source \citep{Steeghs2006}.

The XRT data were processed with standard procedures ({\tt xrtpipeline} v0.10.6),
filtering and screening criteria by using FTOOLS in the
{\tt Heasoft} package (v.6.1.2). 
Given the low rate of the source during the whole campaign,  
we only considered photon counting data 
(PC) and further selected XRT grades 0--12 (\citealt{Burrows2005:XRT}). 
With the exception of observation 6 (2006 Feb 9) 
the data show an average count rate of $<0.5$ counts s$^{-1}$ and 
no pile-up correction was necessary. We therefore extracted the source events 
from a circular region with a radius of 11 pixels (1 pixel $\sim2.37$\arcsec).
During observation 6, 
pile-up correction was required and we adopted an annular source extraction 
region with radii 4 and 30 pixels. 
To account for the background, we also extracted events within an 
annular region centered on the source, and with radii 40 and 100 pixels, 
free from background sources. 
Ancillary response files were generated with {\tt xrtmkarf}, 
and account for different extraction regions, vignetting and 
PSF corrections. We used the latest spectral redistribution matrices 
(v008) in the Calibration Database maintained by HEASARC.
For timing analysis, the arrival times of XRT events were
converted to the Solar System barycentre with the task 
{\tt barycorr} and source events were extracted from the circular 
region with 30 pixels radius to maximize statistics.

BAT always observed \src\ simultaneously with XRT, but
only survey data products, in the form of Detector Plane
Histograms (DPH) with typical integration time of 
$\sim 300$\,s, are available.
The BAT data were analysed using the standard BAT analysis 
software distributed within FTOOLS.
DPH data were calibrated with the task {\tt baterebin} using the
proper BAT gain/offset files from the housekeeping data directory,
and sky images of each observation were extracted in the
15--25, 25--50, 50--100, 100--150 and 15--150\,keV energy bands.
The {\tt batcelldetect} task never detected the source above a
signal-to-noise ratio (S/N) threshold of 4. 
This is consistent with the extrapolation at the high energies 
of the XRT data fit (Sect.~\ref{igr112:dataanal}) with an  
absorbed power-law with exponential cutoff,
with e-folding energy of $15\pm2$\,keV drawn from the RXTE fit
\citep{Swank2007:atel999}.
 
Throughout this paper the uncertainties are given at 
90\% confidence level 
for one interesting parameter (i.e., $\Delta \chi^2 =2.71$) 
unless otherwise stated.
The spectral indices are parameterized as  
$F_{\nu} \propto \nu^{-\alpha}$, 
where $F_{\nu}$ (erg cm$^{-2}$ s$^{-1}$ Hz$^{-1}$) is the 
flux density as a function of frequency $\nu$; 
we also use $\Gamma = \alpha +1$ as the photon index, 
$N(E) \propto E^{-\Gamma}$ (ph cm$^{-2}$ s$^{-1}$ keV$^{-1}$). 

  \section{Analysis and Results\label{igr112:dataanal}}

	\begin{figure}[t]
	 	\includegraphics[angle=270,width=9cm]{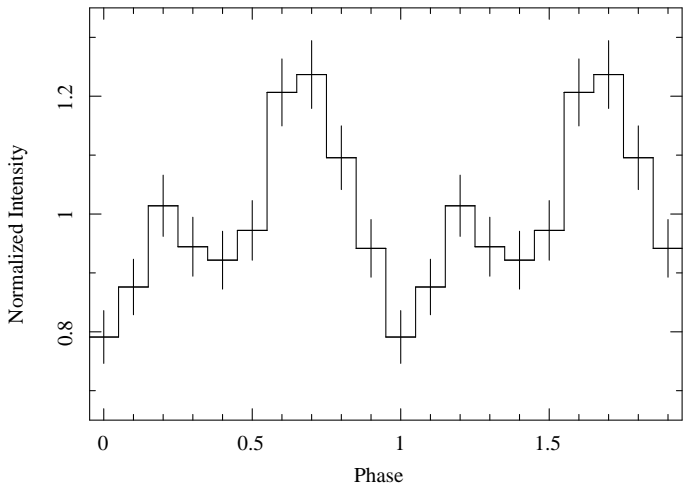}
 		\caption{Folded 0.2--10\,keV light curve of the combined observations 6 though 11, 
		using a period of 186.78\,s \citep{Swank2007:atel999}.
		}
                \label{igr112:fig:foldlcv}
	\end{figure}

A refined position was obtained by summing all 
data taken in 2007 with the exclusion of observation 6  
(affected by pile-up)  
at RA(J2000$)=11^{\rm h} 21^{\rm m} 46\fs90$,   
Dec(J2000$)=-59^{\circ}$ $51^{\prime} 46\farcs9$, with an error, 
drawn from the cross-correlation with the USNO-B1.0 catalogue, 
of $1\farcs1$ (90\% confidence).  
This position is $1\farcs2$ from the optical counterpart HD~306414. 

We extracted light curves in the 1--10\,keV (total), 1--4\,keV (soft) and 4--10\,keV 
(hard) bands. The 0.2--1\,keV band was not used in our analysis because, given the high
absorbing column density, its signal was significantly lower than the one of the
other bands. 
The light curves were corrected for Point-Spread Function 
(PSF) losses, due to the extraction region geometry, 
bad/hot pixels and columns falling within this region, and for vignetting, by using
the task {\tt xrtlccorr} (v0.1.9), which generates 
an orbit-by-orbit correction based on the instrument map. 
We then subtracted the scaled background rate in each band from their 
respective source light curves and calculated the 4--10/1--4 hardness ratio. 
The \src\ light curve (Fig.~\ref{igr112:fig:lcv}) shows an increase in count
rate by a factor of $\sim 10$ in less than 1.5 hours, and of a factor of 
$\sim 65$ in 17 hours on 2007 Feb 9.
However, no significant variation in the hardness ratio can be evidenced
(panels c,d of  Fig.~\ref{igr112:fig:lcv}). 
Indeed, fitting the hardness ratio as a function of time (or as a function 
of count rate) to a constant model yields a 
value of $0.49\pm0.03$ and $\chi^2=1.13$ for 80 degrees of freedom, d.o.f.

We folded the data at the period of 186.78\,s reported by \citet{Swank2007:atel999}
based on Feb 9 01:20--03:20 UT RXTE/PCA observations and obtained the 0.2--10\,keV 
light curve shown in Fig.~\ref{igr112:fig:foldlcv}.

Upon examination of the light curve presented in Fig.~\ref{igr112:fig:lcv} and the 
available counting statistics, we selected different time bins over which we accumulated 
spectra. These include 
{\it i)} the quiescent phase before the 2007 Feb 9 outburst, 
{\it ii)} the Feb 9 outburst (observation 6) and 
{\it iii)} the tail phase of the outburst (observations 7--8, 9--12, 7--12, 7--18). 
We further selected 5 flaring episodes from observation 6 (see Fig.~\ref{igr112:fig:lcv}b). 
A comparison with the data collected during the tail of the 2006 outburst was also performed. 
The data were rebinned with a minimum of 
20 counts per energy bin to allow $\chi^2$ fitting. 
However, for the 2006 observation performed (28 counts),  
before the onset of the outburst (41 counts), 
for the late flares, and the late observations, the Cash statistic 
\citep{Cash1979} and spectrally unbinned data were used. 
The spectra were all fit with XSPEC (v11.3.2) in the 0.5--9\,keV energy range,
adopting the typical pulsar spectral model, an absorbed power law model.

	\begin{figure}[t]
	 	\includegraphics[angle=270,width=9cm]{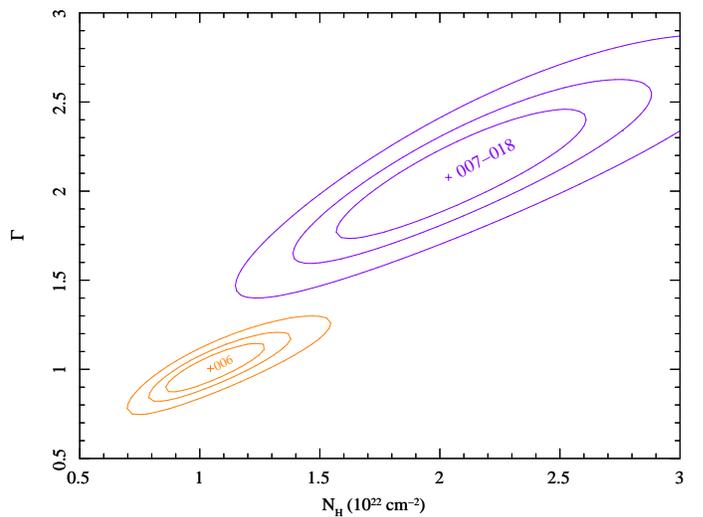}
 		\caption{XRT time-selected spectroscopy. 
		The $\Delta \chi^2 =2.3, 4.61, 9.21$ contour levels for the 
		column density in units of $10^{22}$ cm$^{-2}$  
		vs.\ the photon index, with best-fit values indicated by crosses. 
		Orange contours are from observation 6, while the blue
		ones are from the combined observations 7 though 18. 
		}
                \label{igr112:fig:spec}
	\end{figure}

The best fit parameters are reported in Table~\ref{igr112:tab:specfits} along with the mean 
luminosity of each time selection. 
The spectrum of the brightest part of the outburst (observation 6) could be fit with a 
single power law, with a photon index $\Gamma=1.00_{-0.14}^{+0.16}$ and an absorbing column 
density of $N_{\rm H}=(1.04_{-0.20}^{+0.25})\times 10^{22}$ cm$^{-2}$ 
($\chi^2_{\rm red}=1.04/83$ d.o.f.), while the combined observations 7--18 yielded 
a photon index of $2.08_{-0.37}^{+0.41}$ and 
$N_{\rm H}=(2.04_{-0.50}^{+0.62})\times 10^{22}$ cm$^{-2}$ ($\chi^2_{\rm red}=1.19/19$ d.o.f).
We note that the 1--10\,keV count rate to unabsorbed 1--10\,keV flux conversion factor, 
obtained from the best fit model for observation 6, is  
$2.9 \times 10^{-10}$ erg cm$^{-2}$ count$^{-1}$. 
We also performed fits of observation 6 with an absorbed black-body.
Since the column density assumed a value significantly below that resulting 
from the interstellar medium, the power law fit was favoured.
In all cases the best-fit absorbing column density is consistent (within 2-$\sigma$) 
with the Galactic absorption along the line of sight of \src. 
This value is significantly lower than the column density measured with RXTE/PCA during
the 2006 outburst \citep{Smith2006a}, $(11\pm3)\times$10$^{22}$~cm$^{-2}$, which is likely
overestimated because it was derived with RXTE/PCA in the energy range 2.5--15~keV.

To investigate spectral variations, we created the 
contour levels for the column density vs.\ the photon index. The most interesting 
example is shown in Fig.~\ref{igr112:fig:spec}. 
They indicate that the photon index showed significant variations, 
with the spectrum softening as the outburst progresses, 
confirming the 2006 observation of the outburst tail; 
there is also evidence of an increasing absorbing column density. 
These findings are independent on our choice of an absorbed power-law model.

\section{Discussion and Conclusions\label{igr112:discussion}}

We have carried out the most complete monitoring campaign of an outburst 
from a SFXT, thanks to the known
periodicity of the outburst activity  from \src\ (Paper I).
This is remarkable, since the transient and unpredictable nature 
of the outbursts from all other SFXTs hampers a similar extensive study, 
from the almost ``quiescent'' level up to the ``flaring'' activity.
The entire ``outburst event'' was monitored for 23 days. 
The source was under the threshold of detectability in the early days 
of the campaign, with a luminosity  below  3.7$\times 10^{33}$ erg s$^{-1}$.
On Feb 9 the source underwent a bright outburst up to 
$\sim 1.1 \times 10^{36}$~erg~s$^{-1}$.
The bright part of the outburst (Feb 9; see Fig.~\ref{igr112:fig:lcv}b)
is composed of at least five flares, with variable peak flux,
each with a duration of $\sim15$~min--2~hours. 
This bright flaring activity lasted about 1~day, 
then the source underwent a decline phase, not flat,
but composed of other equally short flares, one order of magnitude fainter.
This decline phase lasted about 5 days, and then the source faded to a much fainter
level, almost below the threshold of detectability.
The whole outburst lasted about 15\,d,	
after which the source became fainter than 1.2 $\times 10^{33}$ erg s$^{-1}$. 
Thus, \src\ reached the typical luminosity of SFXTs during outburst 
(around $10^{36}$~erg~s$^{-1}$), showing a dynamic range larger than $10^3$ 
and a hard X--ray spectrum, proper of this kind of sources. 
The brightest part of the outburst
lasted less than a day, on 9th February, and would have been 
the only flaring activity seen with less sensitive instruments.
Indeed up to now, observations of outbursts from SFXTs have been mostly 
performed with instruments on-board RXTE and/or INTEGRAL, which could only  
catch the brightest flares, and missed the complete evolution 
of the phenomenon, from the onset of the outburst, and down again 
to the level of almost quiescence, which is expected at a level around 
$\sim10^{32}$~erg s$^{-1}$ (possible magnetospheric emission plus 
the contribution from the soft X--ray emission from the OB supergiant). 
Only during XMM-Newton and Chandra
observations of IGR~J17544-2619 \citep{Gonzalez2004,intZand2005}
 outbursts were observed starting from the quiescent emission, 
but the ``post-flare'' phases could  not be completely followed 
and thus the duration of the entire outburst phase could not 
be measured.

For \src\ we could exceptionally observe the whole phenomenon, 
which  for the first time reveals that the ``short outbursts'' 
(the ``flares'' lasting minutes or few hours), are actually part 
of a much longer  ``outburst event'' (lasting several days), which
we believe is triggered at the periastron passage in a wide, highly eccentric orbit. 
Indeed, the \src\ outburst recurrence time (329 days) is remarkably 
stable and reveals an underlying clock, which can be 
naturally associated with the orbital motion in a non-circular orbit.
The short flares most prominent on Feb 9 are 
probably produced by the episodic accretion of clumps 
from the massive wind \citep{Owocki2006},
or by an inhomogeneous accretion stream near periastron 
(similar to what proposed to explain the periodic outbursts from the 
eccentric X-ray pulsar GX~302-1, e.g. \citealt{Leahy1991}).
Thus, both mechanisms originally proposed to explain the SFXTs outbursts, 
seem to be at work in \src, i.e., accretion at periastron passage in a wide
eccentric orbit \citep{Negueruela2005a}, and accretion from clumpy winds, 
\citep{intZand2005}. 
Applying a spherically symmetric homogeneous wind model to 
a B1 Ia spectral type companion, with a mass of 39~$M_\odot$,  
42 solar radii, and a wind mass loss of 3.67$\times$10$^{-6}$ ~$M_\odot$~yrs$^{-1}$  
\citep{Vink2000}, the short outburst duration implies an eccentricity 
larger than 0.9.
From the spectroscopy of the single flares there is evidence for
only minor variations in the local absorbing column density
(which would suggest the clear presence of clumps). 
This may be partly due to the high column density along the line of sight
that absorbs most of the radiation below 1\,keV, thus preventing us from
detecting comparatively small variations of the intrinsic column density.                       
However, XRT data show evidence of softening of the spectrum in the
long decay to quiescent state (thus confirming the 2006 observation
of the outburst tail) and a possible evidence of an $N_{\rm H}$ growth
connected with the same transition.

 \begin{table*}
 \begin{center}
 \caption{Observation log.}
 \label{igr112:tab:alldata}
 \begin{tabular}{lllll}
 \hline
 \hline
 \noalign{\smallskip}
Sequence         & Start time (MJD)     & Start time  (UT)             & End time   (UT)                 & Net Exposure$^{\mathrm{a}}$    \\
                 &              	& (yyyy-mm-dd hh:mm:ss)           & (yyyy-mm-dd hh:mm:ss)        &(s)       \\
 \noalign{\smallskip}
 \hline
 \noalign{\smallskip}
00030384001     &       53814.7853    &	    2006-03-20 18:50:47     &	    2006-03-20 19:01:53     &	    643     \\
00030881001     &       54135.5060    &	    2007-02-04 12:08:34     &	    2007-02-04 23:38:58     &	    2048    \\
00030881002     &       54136.5092    &	    2007-02-05 12:13:12     &	    2007-02-05 23:44:57     &	    1865    \\
00030881003     &       54137.1798    &	    2007-02-06 04:18:55     &	    2007-02-06 17:28:17     &	    1941    \\
00030881004     &       54138.1244    &	    2007-02-07 02:59:06     &	    2007-02-07 17:13:56     &	    1213    \\
00030881005     &       54139.0604    &	    2007-02-08 01:27:02     &	    2007-02-08 16:03:57     &	    1403    \\
00030881006     &       54140.0021    &	    2007-02-09 00:03:05     &	    2007-02-09 23:59:57     &	    4668    \\
00030881007     &       54141.6747    &	    2007-02-10 16:11:33     &	    2007-02-11 00:16:56     &	    3141    \\
00030881008     &       54142.0696    &	    2007-02-11 01:40:15     &	    2007-02-12 00:10:11     &	    4232    \\
00030881009     &       54143.6078    &	    2007-02-12 14:35:17     &	    2007-02-12 19:38:57     &	    3337    \\
00030881010     &       54144.0085    &	    2007-02-13 00:12:17     &	    2007-02-13 16:30:58     &	    3091    \\
00030881011     &       54145.0182    &	    2007-02-14 00:26:09     &	    2007-02-14 21:25:56     &	    4521    \\
00030881012     &       54146.0142    &	    2007-02-15 00:20:26     &	    2007-02-15 09:58:57     &	    4590    \\
00030881013     &       54147.6084    &	    2007-02-16 14:36:05     &	    2007-02-16 19:44:56     &	    5230    \\
00030881014     &       54148.6848    &	    2007-02-17 16:26:09     &	    2007-02-17 21:23:57     &	    4295    \\
00030881015     &       54149.2812    &	    2007-02-18 06:44:59     &	    2007-02-18 12:01:58     &	    4804    \\
00030881016     &       54150.2187    &	    2007-02-19 05:14:52     &	    2007-02-19 12:02:56     &	    4636    \\
00030881017     &       54151.6337    &	    2007-02-20 15:12:28     &	    2007-02-20 20:19:56     &	    4847    \\
00030881018     &       54152.0412    &	    2007-02-21 00:59:16     &	    2007-02-21 17:11:57     &	    5194    \\
00030881019 	&	54153.4431    &	    2007-02-22 10:38:04     &	    2007-02-22 15:39:58     &	    3814    \\
00030881021     &       54155.1683    &	    2007-02-24 04:02:20     &	    2007-02-24 13:46:57     &	    2963    \\
00030881023     &       54157.5108    &	    2007-02-26 12:15:37     &	    2007-02-26 18:48:58     &	    786     \\
 \noalign{\smallskip}
  \hline
  \end{tabular}
  \end{center}
  \begin{list}{}{} 
  \item[$^{\mathrm{a}}$] The exposure time is spread over several snapshots (single continuous pointings at the target)
	during each observation.
  \end{list}   
\end{table*}

 \setcounter{table}{1}	
\begin{table} 	
 \begin{center} 	
 \caption{Spectral fit results.} 	
 \label{igr112:tab:specfits}
 \begin{tabular}{lllrrr} 
 \hline 
 \hline 
 \noalign{\smallskip} 
 Spectrum$^{\mathrm{a}}$ & $N_{\rm H}$           & $\Gamma$              & $\chi^2$ (d.o.f.)/ & 	$L_{1 - 10 {\rm \,keV}}$$^{\mathrm{b}}$\\ 
 		& (10$^{22}$ cm$^{-2}$)	         & 	                  & C-stat(\%)$^{\mathrm{c}}$  &    (erg s$^{-1}$) 	     \\
 \noalign{\smallskip} 
 \hline 
 \noalign{\smallskip} 
001 (2006)	   &	$0.88_{-0.62}^{+0.96}$	&	$1.89_{-0.92}^{+1.07}$	  & 167.7 (65.8\%)  &$2.35\times10^{-1}$     \\
001--005   &	$2.28_{-1.50}^{+2.21}$	&	$1.34_{-0.96}^{+1.11}$	  & 225.9 (65.8\%)  &$4.32\times10^{-2}$     \\
006	   &	$1.04_{-0.20}^{+0.25}$	&	$1.00_{-0.14}^{+0.16}$	  &1.04 (83)  	    &$4.78$     \\
006 Flare 1&	$0.85_{-0.32}^{+0.46}$	&	$0.94_{-0.28}^{+0.31}$	  &0.66 (17) 	    &$8.68$      \\
006 Flare 2&	$1.11_{-0.49}^{+0.79}$	&	$0.91_{-0.32}^{+0.42}$	  &1.16 (25)	    &$8.32$      \\
006 Flare 3&	$0.83_{-0.42}^{+0.62}$	&	$0.82_{-0.40}^{+0.44}$	  &0.93 (10)	    &$11.1$      \\
006 Flare 4&	$0.88_{-0.31}^{+0.38}$	&	$1.03_{-0.31}^{+0.32}$	  &623.6 (39.0\%)   &$4.86$      \\
006 Flare 5&	$2.02_{-0.79}^{+1.01}$	&	$1.51_{-0.53}^{+0.58}$	  &436.0 (54.6\%)   &$2.75$      \\
007--008  &	$1.75_{-0.83}^{+1.03}$	&	$1.94_{-0.60}^{+0.64}$	  &518.3 (66.0\%)   &$2.22\times10^{-1}$      \\
009--012  &	$1.04_{-0.36}^{+0.47}$	&	$1.48_{-0.35}^{+0.39}$	  &576.9 (53.6\%)   &$8.82\times10^{-2}$      \\
007--012  &	$1.86_{-0.44}^{+0.68}$	&	$1.92_{-0.38}^{+0.43}$	  &1.09 (18)	    &$1.40\times10^{-1}$      \\
007--018  &	$2.04_{-0.50}^{+0.62}$	&       $2.08_{-0.37}^{+0.41}$	  &1.19 (19)	    &$5.96\times10^{-2}$	\\
  \noalign{\smallskip}
  \hline
  \end{tabular}
  \end{center}
  \begin{list}{}{} 
  \item[$^{\mathrm{a}}$] Last three digits of observation numbers, see Table~\ref{igr112:tab:alldata}, column 1. 
  \item[$^{\mathrm{b}}$] Luminosity in the 1--10\,keV band in units of $10^{35}$ erg s$^{-1}$ obtained from the spectral fits. 
  \item[$^{\mathrm{c}}$] Cash statistics (C-stat) and percentage of Monte Carlo realizations that had statistic $<$ C-stat. 
			We performed $10^4$ simulations.
  \end{list} 
  \end{table} 

\begin{acknowledgements}

We thank the Swift team for making these observations possible,
in particular the duty scientists and science planners 
M.\ Chester, S.\ Hunsberger, J.\ Kennea, C.\ Pagani and J.\ Racusin; 
we thank N.\ Gehrels for approving this ToO and D.\ Burrows for a winning observing strategy. 
We thank S.\ Campana, P.\ D'Avanzo, A.\ Paizis,  P.\ Persi, V.F.\ Polcaro, and S.\ Vercellone 
for insightful discussions. 
This research has made use of NASA's Astrophysics Data System Bibliographic Services 
as well as the NASA/IPAC Extragalactic Database (NED) which is operated 
by the Jet Propulsion Laboratory, California Institute of Technology, under contract with 
the National Aeronautics and Space Administration. 
This work was supported by MIUR grant 2005-025417, and contract ASI/INAF I/023/05/0.
PR thanks INAF-IASFMi, where most of the work was carried out, for their kind hospitality. 
\end{acknowledgements}

\end{document}